\documentclass[aps,prl,twocolumn,groupedaddress,showpacs]{revtex4}

\usepackage{graphicx}
\usepackage{dcolumn}
\usepackage{bm}

\def\bc{\begin{center}}
\def\ec{\end{center}}

\def\beq{\begin{equation}}
\def\eeq{\end{equation}}

\def\sm{\sigma_{xx}^{\min}}

\begin{document}


\title{Robust Transport Properties in Graphene}

\author{K. Ziegler}%
\email{Klaus.Ziegler@Physik.Uni-Augsburg.de}
\affiliation{
Institut f\"ur Physik, Universit\"at Augsburg, D-86135 Augsburg, Germany
}

\date{\today}

\begin{abstract}
Two-dimensional Dirac fermions are used to discuss quasiparticles
in graphene in the presence of impurity scattering. Transport properties
are completely dominated by diffusion. This may explain why recent 
experiments did
not find weak localization in graphene. The diffusion coefficient of the 
quasiparticles decreases strongly with increasing strength of disorder. 
Using the Kubo formalism, however, we find a robust minimal conductivity
that is independent of disorder. This is a consequence of the fact that
the change of the diffusion coefficient is fully compensated by a change
of the number of delocalized quasiparticle states.    
\end{abstract}

\pacs{81.05.Uw, 71.55.Ak, 72.10.Bg, 73.20.Jc}
\maketitle






\maketitle


Recent experimental studies of single graphite layers (graphene) have 
revealed interesting transport properties 
\cite{novoselov05,zhang05,morozov06}.
A quantum Hall effect was found in the presence of an external magnetic 
field with Hall plateaus $\sigma_{xy}=\pm(2n+1)2e^2/h$ 
($n=0,1,...$). This result can be explained by the 
band structure of graphene which has two nodes (or valleys) due to the 
hexagonal lattice and a linear dispersion around each of these 
nodes \cite{semenoff84}. 
If the Fermi energy is near these nodes the quasiparticles can be
described as fourfold degenerated (two valleys and two spins orientations) 
2D Dirac fermions. Another interesting observation is the
existence of a minimal conductivity $\sm$ which occurs
if the Fermi energy is exactly at the nodes. This quantity shows a
remarkable stability with $\sm=3...5$ $e^2/h$, even if the
mobility of the studied samples changes by almost a factor 6 from 
$\mu = 0.15 m^2/Vs$ to $\mu = 0.85 m^2/Vs$ \cite{novoselov05}.
The varying mobilities in different samples indicate a
varying density of impurities. The transport mechanism must
be related to diffusion of electrons and holes, caused by impurity 
scattering. The latter can formally be described by a random 
potential. Then, in terms of
the Dirac fermions, the impurities can create gap fluctuations,
i.e. a random Dirac mass, because a gap can easily be opened at
the nodes of the band structure, for instance, by a staggered
potential \cite{haldane88}. 

Existing theories of 2D Dirac fermions predicted the value 
$\sigma_{xx}=e^2/h\pi$ for a single Dirac node 
\cite{fradkin86,lee93,ludwig94,ziegler98,peres06}, 
(i.e. $\sm=4e^2/h\pi$) such that there is a quantitative discrepancy 
by a factor $1/\pi$ in comparison with 
the experimental observation, as discussed in Ref. 
\cite{novoselov05}. In recent papers several authors applied
the Landauer formula, instead of the Kubo formula used in previous studies, 
to determine the minimal conductivity also
as $\sm=4e^2/h\pi$ for a rectangular system with
aspect ratio $W/L\gg 1$ \cite{katsnelson05,tworzydlo06}.
Nomura and MacDonald argued that $\sm$ could be enhanced by
Coulomb scattering, leading to $\sm=4e^2/h\pi$ \cite{nomura06}.
Several possibilities for the value of $\sm$, using different
approaches to the linear response were also 
discussed in Ref. \cite{cserti06}.  

The effect of quantum interference due to impurity scattering
was studied in a recent experiment \cite{morozov06}. It was found 
that there is no weak localization in a single graphene sheet. 
Multilayer graphite films, on the other hand, exhibit clearly weak 
localization. These observations indicate that graphene has special 
transport properties.

It will be shown in the following that (1) a calculation, based 
on linear response theory (Kubo formula), gives a conductivity of
$\sm=\pi e^2/h$ for the pure system, 
(2) weak scattering leads to a linear Boltzmann conductivity similar
to what was observed experimentally, and
(3) there are no weak (anti-)localization corrections due to a
spontaneously broken supersymmetry which creates diffusive fermions. 


Starting from the Kubo formula \cite{madelung}, 
the conductivity tensor $\sigma_{\mu\nu}$ of a system with 
Hamiltonian $H$ at inverse temperature
$\beta=1/k_BT$ and for frequency $\omega$ reads
\begin{equation}
{e\over i\hbar}\lim_{\alpha\to0}\int_{-\infty}^0
e^{(i\omega+\alpha)t}Tr\Big([e^{-\beta H},r_\mu]e^{-iHt}j_\nu 
e^{iHt}\Big)dt .
\label{kubo0}
\end{equation}
The current operator is given by the Hamiltonian 
as the commutator
$
j_\nu=-ie[H,r_\nu]
$.
For non-interacting fermions with single-particle
energy eigenstates 
$|k\rangle$ (i.e. $H|k\rangle=\epsilon_k|k\rangle$)
a lengthy but straightforward calculation yields
\begin{eqnarray}
\sigma_{\mu\nu}=-i{e^2\over\hbar}
\sum_{k_1,k_2}\langle k_1|[H,r_\mu]|k_2\rangle
\langle k_2|r_\nu|k_1\rangle
\nonumber \\
\times{f_\beta(\epsilon_{k_2})-f_\beta(\epsilon_{k_1})
\over \epsilon_{k_1}-\epsilon_{k_2}+\omega-i\alpha}
\label{kubo1}
\end{eqnarray}
with Fermi function $f_\beta(\epsilon)=1/(1+e^{-\beta\epsilon})$.
The identity
$
\langle k_2|[H,r_\nu]|k_1\rangle
=(\epsilon_{k_2}-\epsilon_{k_1})\langle k_2|r_\nu|k_1\rangle
$
and the Dirac delta function $\pi\delta(\epsilon_k-\epsilon)
=\lim_{\eta\to0}\eta/((\epsilon_k-\epsilon)^2+\eta^2)$
allow us to express the conductivity 
in Eq. (\ref{kubo1}) as a double integral with respect to two energies 
$\epsilon, \epsilon'$:
\[
\sigma_{\mu\nu}=i{e^2\over\hbar}
\int\int Tr\Big\{[H,r_\mu]
\delta(H-\epsilon')[H,r_\nu]\delta(H-\epsilon)
\Big\}
\]
\beq
\times
{1\over \epsilon-\epsilon'+\omega-i\alpha}
{f_\beta(\epsilon')-f_\beta(\epsilon)
\over\epsilon-\epsilon'}d\epsilon d\epsilon' .
\label{kubo2}
\eeq
An alternative expression is obtained by writing the Dirac delta 
functions in terms of the Green's function $G(z)=(H-z)^{-1}$. 
Using the identity
\[
G(z)[H,r_\mu]G(z')=r_\mu G(z')-G(z)r_\mu+G(z)r_\mu(z-z')G(z') ,
\]
the diagonal elements of the conductivity tensor read
\begin{eqnarray}
\sigma_{\mu\mu}=
i{e^2\over\hbar}{1\over8\pi^2}\lim_{\eta_1,\eta_2\to0}\int\int 
\sum_{r,r'}(r_\mu-r_\mu')^2
\sum_{s_1,s_2=\pm1}s_1s_2
\nonumber \\
(\epsilon'-\epsilon+i(s_1\eta_1-s_2\eta_2))
Tr_n[G_{rr'}(\epsilon'+is_1\eta_1)
\nonumber \\
G_{r'r}(\epsilon+is_2\eta_2)]
{f_\beta(\epsilon')-f_\beta(\epsilon)
\over \epsilon-\epsilon'+\omega-i\alpha}
d\epsilon d\epsilon' ,
\label{kubo3}
\end{eqnarray}
where $Tr_n$ is the trace related to $n$ additional 
degrees of freedom (e.g., $n=2$ for the spinor index of 2D Dirac 
fermions).

In the special case of graphene the Hamiltonian reads in sublattice
representation  \cite{semenoff84,haldane88}
\beq
H=\sigma_1h_1+\sigma_2h_2 ,
\label{hamiltonian}
\eeq
where $\sigma_j$ ($j=1,2,3$) are Pauli matrices. 
In Fourier representation with wavevector $\vec{k}=(k_1,k_2)$ 
the coefficients of the Pauli matrices in a pure system are
\[
h_1=-t\sum_{j=1}^3\cos(\vec{a}_j\cdot\vec{k}) ,
\hskip0.5cm
h_2=-t\sum_{j=1}^3\sin(\vec{a}_j\cdot\vec{k})
\] 
with the lattice vectors of the honeycomb lattice
$\vec{a}_{1}=(-\sqrt{3}/2,1/2)$, $\vec{a}_{2}=(0,-1)$ and
$\vec{a}_{3}=(\sqrt{3}/2,1/2)$. $H$ can be diagonalized 
as $H=diag(e_k,-e_k)$ with $e_k=\sqrt{h^2_1+h^2_2}$.
The current operator transforms under Fourier transformation as
$-ie[H,r_\mu]\to e\partial H/\partial k_\mu$.

There are six points at 
$\vec{k}=(\pm 4\pi/3\sqrt{3},0)$, $(2\pi/3\sqrt{3},\pm 2\pi/3)$,
and $(-2\pi/3\sqrt{3},\pm 2\pi/3)$ where $e_k$ vanishes, corresponding
with the two nodes. The commutators are in the diagonal representation 
of $H$
\beq
[H,r_\mu]_{12}[H,r_\mu]_{21}={1\over e_k^2}
\Big(h_2{\partial h_1\over\partial k_\mu}-h_1{\partial h_2\over\partial k_\mu}\Big)^2
\label{comm}
\eeq
whose value is $9/4$ at all nodes. The 2D $k$ integration at each node
can be expressed by an integration with respect to $h_1$ and $h_2$ as
$
d^2k=J dh_1 dh_2 
$, where the Jacobian is $J=4/9$ at all nodes.
Therefore, after the angular integration around the nodes the integration
is given by 
\[
J(2\pi/3) e_kd e_k=(8\pi/27) e_kd e_k
\hskip0.5cm
(0\le e_k\le\lambda) .
\]
This can be inserted in the conductivity of Eq. (\ref{kubo2})
and after the summation over all nodes the conductivity 
$\sigma_{\mu\mu}$ reads at low temperatures ($\beta\sim\infty$)
\[
-i{e^2\over h}{12\over 27}
\int_0^\lambda\Big\{
{[H,r_\mu]_{12}[H,r_\nu]_{21}\over2e_k+\omega-i\alpha}
+{[H,r_\mu]_{21}[H,r_\nu]_{12}\over -2e_k+\omega-i\alpha}
\Big\} d e_k .
\]
Inserting the commutators from Eq. (\ref{comm}) yields eventually
\beq
Re(\sigma_{22})={e^2\over h}\int_0^\lambda\pi
\delta(2e_k-\omega)d e_k ={\pi\over 2}{e^2\over h} .
\label{cond1}
\eeq
Another factor of 2 comes from the spin-1/2 degeneracy of
the quasiparticles. Thus our calculation gives for the minimal 
conductivity $\sm=\pi e^2/h$.

\begin{figure}
\centering
\includegraphics[width=0.45\textwidth]{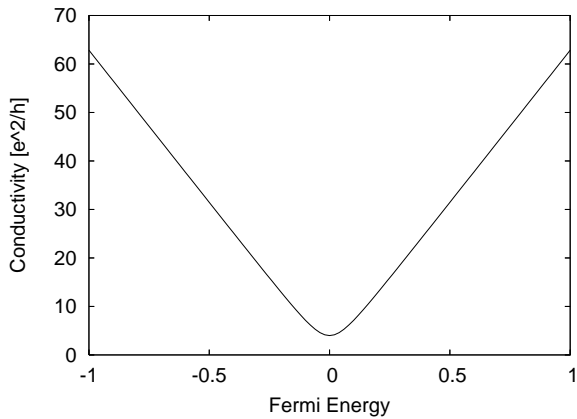}
\caption{Conductivity of graphene calculated in mean-field approximation 
(from Eq. (\ref{cond5})).}
\label{plotcond}
\end{figure}

In a pure graphene sheet there is only ballistic transport. Consequently,
the diffusion coefficient $D$ is infinite. On the other hand, 
if the Fermi energy is exactly at the nodes, the related density of 
states $\rho$ vanishes. From this point of view, the conductivity, 
expressed by the Einstein relation as $\sm\propto\rho D$,  depends very 
sensitively on the limits of the model parameters (e.g., the DC limit
$\omega\to0$). A more instructive situation is a system with
randomly distributed scatterers that may lead to diffusion (i.e. $D<\infty$)
or even to Anderson localization (i.e. $D=0$) 
\cite{anderson58,abrahams79,hatsugai93,pereira06}. 
A source of disorder in the 
tight-binding Hamiltonian $H$ of Eq. (\ref{hamiltonian}) is a randomly 
fluctuating nearest-neighbor hopping rate.
For a qualitative discussion of random scattering, the Hamiltonian
is approximated by the 2D Dirac Hamiltonian
\[
H_D=i\sigma_1\nabla_1+i\sigma_2\nabla_2+m\sigma_3 .
\]
A randomly fluctuating gap is introduced by a random Dirac mass
$m$ with Gaussian distribution of zero mean and variance $g$.
The transformation property of the tight-binding Hamiltonian
\beq
H^T=-\sigma_2 H \sigma_2 
\label{transf1}
\eeq
is also obeyed by the random Dirac Hamiltonian $H_D$ after rotating
$\sigma_1\to\sigma_2$ and $\sigma_2\to\sigma_1$.
This property is crucial for the formation of a diffusion mode
in two dimensions \cite{ziegler98}. Models which violate this property, 
e.g. by an additional term proportional to a $2\times2$ unit matrix,
may lead to localization of states near the Fermi energy 
\cite{hatsugai93,pereira06}.  
Intervalley scattering is ignored by the approximation $H\approx H_D$ 
such that we have only independent 
Dirac cones. The effect of the random mass can be studied by applying a
perturbation theory, using a partial summation of 
an infinite series of 
most relevant 
contributions. On the level of the averaged single-particle Green's function
this leads to a selfenergy term
$\eta$ in the Green's function: $G_\pm(i\epsilon)\equiv G(i\epsilon\pm i\eta)$.
This is formally equivalent to a mean-field approximation of a supersymmetric 
functional-integral approach, where the random Dirac mass is replaced 
by a random supermatrix \cite{ziegler98}. Then $\eta$ is obtained as a 
solution of
\beq
\eta=g(\eta+\epsilon)
\int[(\eta+\epsilon)^2+k^2]^{-1}k dk/\pi .
\label{SP}
\eeq
The Green's function can be expressed again by a perturbation expansion,
now in terms of fluctuations around the mean-field approximation 
$G_\pm(i\epsilon)$. 
This expansion can be inserted into the conductivity (\ref{kubo2}). 
The leading term for $i\epsilon=E_F$ (Fermi energy) is the Boltzmann
conductivity
\begin{eqnarray}
Re(\sigma_{\mu\mu})\approx
{e^2{\bar\sigma}\eta^2\over\hbar\pi}
\int \Big[
((e_k-E_F)^2+\eta^2)^{-2}
\nonumber \\
+((e_k+E_F)^2+\eta^2)^{-2}
\Big]\rho(e_k)de_k .
\label{cond5}
\end{eqnarray}
Here ${\bar\sigma}$ is an approximation of the two commutators in Eq.
(\ref{kubo2}) and $\rho$ is the density of states of pure Dirac fermions:
$\rho(E)=\rho_0|E|$, where $\rho_0$ depends on the cutoff $\lambda$ 
of the spectrum of $H$. 
This is a classical result for the conductivity 
of Dirac fermions that was already anticipated by Fradkin 
\cite{fradkin86} and discussed in
the context of d-wave superconductors by Lee \cite{lee93}
and for graphene by Peres et al. \cite{peres06}.
The conductivity at $E_F=0$ is $\sm\approx2e^2{\bar\sigma}\rho_0/h$ 
and does not depend on $\eta$. Thus $\sm$ is independent of
the strength of impurity scattering $g$. Sufficiently away from
$E_F=0$ the conductivity becomes linear, as shown 
in Fig. \ref{plotcond}. This behavior agrees well with 
the experimentally 
observed linear conductivity \cite{novoselov05}. 

The next question is whether 
or not quantum interference effects are important. 
The corresponding corrections in the conductivity are given by
the next order terms of the expanded Green's function.
This includes a logarithmic term (Cooperons) due to
a massless mode of the fluctuations around the 
mean-field approximation. Previous studies 
found that the corrections 
give an antilocalization effect for conventional scatterers, i.e. an
increase of the conductivity due to quantum interference effects
\cite{suzuura}. Additional terms in the Hamiltonian of second order 
in the momentum can suppress weak antilocalization 
\cite{mccann06}.

A crucial question, implied by the antilocalization effect and
the absence of weak-localized corrections in experiments
\cite{morozov06}, is whether or not the states in graphene are 
localized. The conductivity in Eq. (\ref{kubo2}) does not 
directly address localization, in contrast to 
the alternative expression in Eq. (\ref{kubo3}): The 
long-distance behavior of the Green's functions is directly 
related to the behavior of the quantum states. 
For this purpose we return to Eq. (\ref{kubo3}), consider
the minimal conductivity (i.e. $E_F=0$), 
and take the limits $\alpha\to 0$ and $\beta\to\infty$ 
($\eta_1,\eta_2\to0$ are implicit) to obtain
\begin{eqnarray}
\sm
=-{e^2\over\hbar}{\omega\over8\pi}
\int_{-\omega/2}^{\omega/2}
\sum_{r,r'}(r_\mu-r_\mu')^2\sum_{s_1,s_2=\pm1}s_1s_2
\nonumber \\
Tr_2[G_{rr'}(\epsilon+\omega/2+is_1\eta_1)
G_{r'r}(\epsilon-\omega/2+is_2\eta_2)]
d\epsilon .
\label{conda}
\end{eqnarray}
Assuming that the integrand is finite for $0<\omega\ll 1$ it 
can be pulled out of the integral for $\epsilon=0$. The major
contribution comes from $s_1\ne s_2$, where the poles of the Green's
function are on different sides of the real axis.
A finite factor $\sigma'$ takes care of a correction 
in comparison with the exact value of the integral: 
\begin{eqnarray}
\sm\approx
{e^2\over\hbar}{\sigma'\over2\pi}
\omega^2\sum_{r,r'}(r_\mu-r_\mu')^2
Tr_2[G_{rr'}(\omega/2+i\eta_1)
\nonumber \\
G_{r'r}(-\omega/2-i\eta_1)] .
\label{kubo5}
\end{eqnarray}
For localized states the sum,
averaged with respect to randomness,
is finite due to the exponential decay for $|r-r'|\gg1$.
In the DC limit $\omega\to0$ this would lead to a vanishing $\sm$. 

In order to evaluate the expression in Eq. (\ref{kubo5}) 
it is convenient to return to the functional
integral of Ref. \cite{ziegler98}. It was found that the 
underlying supersymmetry of the integral is spontaneously 
broken by the mean-field solution of Eq. (\ref{SP}). 
The main consequence of this effect is a
diffusive fermionic mode, similar to the Goldstone mode in
systems with a rotational symmetry, that can be 
formally described by a complex Grassmann field $\Psi_{r}$. 
This allows us to write
\beq
\langle
Tr_2[G_{rr'}(i\epsilon)G_{r'r}(-i\epsilon)]\rangle
={-4\eta^2\over g^2}\int {\bar\Psi}_{r}\Psi_{r'}
e^{-S''}{\cal D}[\psi] ,
\label{corr2}
\eeq
where the action depends on the solution $\eta$ of Eq. (\ref{SP}):
\beq
S''={4\eta^2\over g(\eta+\epsilon)}\int[
\epsilon+{g\over4\pi(\eta+\epsilon)}k^2]
{\bar\Psi}_k\Psi_{-k} d^2k .
\label{massless}
\eeq
The latter contains the diffusion coefficient 
\beq
D={g\over4\pi(\eta+\epsilon)} 
\label{dcoeff}
\eeq
that depends strongly on the variance $g$ of the distribution 
of random scatteres.

The minimal conductivity is obtained from Eq. (\ref{corr2})
for small $\epsilon=i\omega/2$, together with Eq. (\ref{kubo5}), as
\beq
\sm=\sigma'{e^2\over h\pi} .
\label{cond6}
\eeq
This agrees with the result of the mean-field approximation
in Eq. (\ref{cond5}), except for the (undetermined) prefactors.
The variation of $\sigma'$ with $g$ for $0\le g\le 1$ can be
neglected. Comparing it with the result in Eq. (\ref{cond1}) 
we conclude that the renormalization factor is $\sigma'=\pi^2$.
Thus, in contrast to the diffusion coefficient $D$ the minimal
conductivity
$\sm$ does not depend on $g$. This implies the absence of
corrections due to quantum interference on characteristic
length scales, in agreement with recent experimental observations
\cite{morozov06}.

The fact that $\sm$ is so robust with respect to 
impurity scattering can be understood in terms of the Einstein 
relation $\sm\propto \rho D$, where the conductivity is
separated into the diffusion coefficient $D$ and the averaged 
density of states $\rho$ at the Fermi energy $E_F=0$. The latter is 
calculated from a functional integral similar to Eq. (\ref{corr2})
as $\rho=\eta/\pi g$ \cite{ziegler98}. Then with $D$ of Eq. (\ref{dcoeff})
the minimal conductivity reads 
\[
\sm\propto \rho D\propto
{\eta\over\eta+\epsilon}={\eta\over\eta+i\omega/2} .
\]
For small $\epsilon$ the mean-field
equation (\ref{SP}) gives $\eta\approx e^{-\pi/g}$ which
implies for $D$ and $\rho$
\beq
D\approx ge^{\pi/g}/4\pi,
\hskip0.3cm
\rho\approx e^{-\pi/g}/\pi g .
\label{dcoeff2}
\eeq
Thus, moving away from the ballistic limit $g=0$, the conductivity
should fall rapidly with increasing random potential fluctuations
due to a decreasing $D$ in the Einstein relation. On the other
hand, the density of states $\rho$ increases correspondingly
so that in $\sm$ the influence of random scattering is compensated.
This is in agreement with the direct evaluation of $\sm$ in Eq. 
(\ref{cond6}). The results for $\rho$ and $D$ in Eq. (\ref{dcoeff2}) 
describe a nonperturbative effect of disorder which is not visible within
an expansion in powers of $g$.

Strong potential scattering by charged impurities in the
substrate, for instance, 
can lead to a destruction of the massless fermion mode
used in Eq. (\ref{massless}). This can cause localization and a 
vanishing $\sm$, at least at very low temperatures. The localized
regime cannot be treated within a conventional field theory but would 
require either a numerical finite-size scaling approach
\cite{mackinnon81} or a strong-disorder expansion.

In conclusion, we have studied the conductivity 
of graphene, using a model of 2D Dirac fermions. In the
case of a pure system the ballistic transport leads to
a minimal conductivity $\sm=\pi e^2/h$.
In the case of impurity scattering we found pure diffusion
for any strength of Gaussian distributed scatterers, where
the diffusion coefficient depends strongly on the distribution.
The minimal conductivity $\sm$, on the other hand, does not
depend on the strength of impurity scattering because the
change of the diffusion coefficient is completely compensated 
by a change of the density of diffusive states. 

\begin{acknowledgments}
This research was supported in part by the Sonderforschungsbereich 484.
\end{acknowledgments}

\end{document}